\documentclass[aps,pre,showpacs,floats,twocolumn]{revtex4}
\usepackage{amssymb}
\usepackage{graphics}
\usepackage[dvipdfm]{graphicx}
\usepackage{epsfig}
\usepackage{dcolumn}
\usepackage{amsmath}

\begin{document}

\title{Thermal transport in a higher-order generalized hydrodynamics}
\author{Cl\'{o}ves G. Rodrigues$^{1}$\footnote{Corresponding author.
\\E-mail addresses: cloves@pucgoias.edu.br (C.G. Rodrigues), bomfim@ita.br (C.A.B.
Silva), jramos@ifi.unicamp.br (J.G. Ramos), rlgmesd@ifi.unicamp.br
(R. Luzzi)}, Carlos A. B. Silva$^{2}$, Jos\'{e} G. Ramos$^{3}$,
Roberto Luzzi$^{3}$} \affiliation{$^{1}$School of Exact Sciences and
Computing, Pontifical Catholic University of Goi\'{a}s, 74605-010,
CP 86, Goi\^{a}nia, Goi\'{a}s, Brazil\\
$^{2}$Departamento de F\'{\i}sica, Instituto Tecnol\'{o}gico de
Aeron\'{a}utica, 12228-901, S\~{a}o Jos\'{e} dos Campos, SP, Brazil\\
$^{3}$Condensed Matter Physics Department, Institute of Physics ``Gleb
Wataghin'' State University of Campinas-Unicamp, 13083-859 Campinas, SP,
Brazil}
\date{\today}

\begin{abstract}
Thermal transport in classical fluids is analyzed in terms of a
Higher-Order Generalized Hydrodynamics (or Mesoscopic
Hydro-Thermodynamics), that is, depending on the evolution of the
energy density and its fluxes of all orders. It is derived in terms
of a Kinetic Theory based on the Non-Equilibrium Statistical
Ensemble Formalism. The general system of coupled evolution
equations is derived. Maxwell times -- which are of large relevance
to determine the character of the motion -- are derived. They also
have a quite important role for the choice of the contraction of
description (limitation in the number of fluxes to be retained) in
the study of the hydrodynamic motion. In a description of order 1 it
is presented an analysis of the technological process of thermal
prototyping.
\end{abstract}

\maketitle


\section{Introduction}

The modern advanced technologies and their resulting end use for
obtaining improved and novel products, creates stress on the basic
sciences of Physics and Chemistry. This is a result of trying to
maintain a balance in the triad ST\&I (Science, Technology and
Innovation). A particular question involves the dissipation of
energy and heat transport in, for example, devices under high-levels
of excitation, namely, working in far-removed-from equilibrium
conditions and eventually involving ultrafast relaxation and
transport processes as well as spatial nanometric scales. Another
important aspect is the one of fluids under flow present in certain
production processes (e.g., in food engineering, petrochemistry,
etc.) whose performance depends on their hydrodynamic properties. We
address here, under the scope of a Higher-Order Generalized
Hydrodynamics (HOGH) the question of heat transport in a fluid
embedded in a thermal bath (i.e., mixed with other fluid acting as
such). On this, it has been noticed that one of the complicated
problems of the non-equilibrium theory of transport processes in
dense gases and liquids is the fact that their kinetics and
hydrodynamics are intimately coupled, and must be treated
simultaneously (e.g., see Refs. [1-3]).

Along the last decades Hydrodynamics has been extensively treated
resorting to the Nonequilibrium Molecular Dynamics (NMD for short).
NMD is a computational method created for modelling physical systems
at the microscopic level, being a good technique to study the
molecular behavior of several physical processes [4,5]. On the other
hand, another very satisfactory approach to deal with hydrodynamics
within an ample scope of non-equilibrium conditions consists in the
kinetic theory based on the Non-Equilibrium Statistical Ensemble
Formalism (NESEF for short) [6-15]. NESEF provides a way to go
beyond standard (or classical Onsagerian) Hydrodynamics which
involves restrictions, namely, local equilibrium; linear relations
between fluxes and thermodynamic forces (meaning weak amplitudes in
the motion with Onsager's symmetry laws holding); near homogeneous
and static movement (meaning that the motion can be well described
with basically Fourier components with long wavelengths and low
frequencies, and then involves only smooth variation in space and
time); and weak and rapidly regressing fluctuations [16,17]. Hence,
more advanced approaches are required to lift these restrictions.
Consider first near homogeneity, which implies validity in the limit
of long wavelengths (or wavenumber $Q$ approaching to zero), and to
go beyond it is necessary to introduce a proper treatment valid, in
principle, for intermediate and short wavelengths (intermediate to
large wavenumbers). In phenomenological theories this corresponds to
go from classical irreversible thermodynamics to extended
irreversible thermodynamics [18-21]. This is what has been called
\emph{generalized hydrodynamics}, a question extensively debated for
decades by the Statistical Mechanics community. Several approaches
have been used, and a description can be consulted in Chapter 6 of
the classical book on the subject by Boon and Yip [22]. Introduction
of nonlocal effects for describing motion with influence of ever
decreasing wavelengths, going towards the very short limit and
involving very high frequencies, has been done in terms of what is
referred to as Higher-Order Generalized Hydrodynamics (HOGH) or
Mesoscopic Hydro-Thermodynamics (MHT) [23,24]. A first construction
based on a large expansion of Extended Irreversible Thermodynamics,
i.e., on phenomenological basis, and centered on the analysis of
heat transport is reported in Ref. [23]. A full construction in the
scope of Non-Equilibrium Statistical Thermodynamics is described in
Refs. [24-26].

We consider here the question of heat transport in classical fluids
within the scope of HOGH. It is obtained a hierarchy of equations of
motion for the density of energy and its fluxes of all orders, which
are all coupled together and including thermo-striction terms that
couple them with the evolution equations of the particle density and
its fluxes of all order. Maxwell-times associated to the different
fluxes are evidenced, which, as already noticed, determine the
characteristics of the motion playing an important role in the
choice of the contraction of description (i.e., to fix the number of
fluxes to be retained) appropriate to the study of the hydrodynamics
motion under consideration.

Finally, resorting to the use of HOGH of order 1 it is described a
full analysis of the physical processes that are involved in a
technological procedure of thermal prototyping.


\section{Theoretical Background}

According to theory, immediately after the open system of $N$
particles, in contact with external sources and reservoirs, has been
driven out of equilibrium, the description of its state requires to
introduce all its observables. But this is equivalent to have access
to the so-called one-particle (or single-particle), $\hat{n}_{1}$,
and two-particle, $\hat{n}_{2}$, dynamical operators for any subset
of the particles involved. This is so because all observable
quantities can be expressed at the microscopic mechanical level in
terms of these operators (e.g. Refs. [27,28]). At the classical
mechanical level used here they are given by
%
\begin{equation}
\widehat{n}_{1}(\mathbf{r},\mathbf{p}|\Gamma) = \sum_{j=1}^{N}\delta
\left( \mathbf{r}-\mathbf{r}_{j}\right) \delta \left( \mathbf{p} -
\mathbf{p}_{j}\right) \, ,
\end{equation}
%
\begin{eqnarray}
\widehat{n}_{2}(\mathbf{r},\mathbf{p};\mathbf{r}^{\prime
},\mathbf{p}^{\prime}|\Gamma) &=& \sum_{j\neq k=1}^{N}\delta \left(
\mathbf{r}-\mathbf{r}_{j}\right) \delta \left(
\mathbf{p}-\mathbf{p}_{j}\right) \times \notag \\
&& \delta \left(\mathbf{r}^{\prime }-\mathbf{r}_{k}\right) \delta
\left(\mathbf{p}^{\prime }-\mathbf{p}_{k}\right) \, .
\end{eqnarray}
In these equations $\mathbf{r}$ and $\mathbf{p}$ are the so called
position and momentum field variables and $\Gamma$ is a point in
phase space of particles at position $\mathbf{r}_{j}$, and with
linear momentum $\mathbf{p}_{j}$.

The non-equilibrium statistical operator, $\varrho_{\varepsilon
}(t)$, is a functional of $\widehat{n}_{1}(\mathbf{r},\mathbf{p})$
and $\widehat{n}_{2}(\mathbf{r},\mathbf{p};\mathbf{r}^{\prime
},\mathbf{p}^{\prime})$, each multiplied by the associated
non-equilibrium thermodynamic variable, which we call
$F_{1}(\mathbf{r},\mathbf{p};t)$ and $F_{2}(\mathbf{r},\mathbf{p};
\mathbf{r}^{\prime},\mathbf{p}^{\prime};t)$ [6,8,9].

The non-equilibrium thermodynamic state [8,9,29,30] associated to
the basic dynamic variables $\widehat{n}_{1}$ and $\widehat{n}_{2}$
are the one-particle and two-particle distribution functions
%
\begin{equation}
f_{1}(\mathbf{r},\mathbf{p};t) = \mathrm{Tr}\{\widehat{n}_{1}(\mathbf{r},%
\mathbf{p}) \varrho_{\varepsilon}\left( t\right) \} \, ,
\end{equation}
%
\begin{equation}
f_{2}(\mathbf{r},\mathbf{p},\mathbf{r}^{\prime },\mathbf{p}^{\prime
};t) =
\mathrm{Tr}\{\widehat{n}_{2}(\mathbf{r},\mathbf{p},\mathbf{r}^{\prime
}, \mathbf{p}^{\prime}) \varrho_{\varepsilon }(t)\}\,,
\end{equation}
that is, the average over the nonequilibrium ensemble of
$\widehat{n}_{1}$ and $\widehat{n}_{2}$. The trace operator is in
this classical approach to be understood as an integration over
phase space; $\widehat{n}_{1}$ and $\widehat{n}_{2}$ are functions
defined on phase space and $\varrho_{\varepsilon}$, as stated, is a
functional of these two. For the sake of completeness we write down
its expression, namely
%
\begin{equation}
\varrho_{\varepsilon}(t) = \exp \Big\{\ln
\bar{\varrho}(t,0)-\int_{-\infty }^{t}dt^{\prime }e^{\varepsilon
(t^{\prime }-t)}\frac{d}{dt^{\prime}} \ln \bar{\varrho}(t^{\prime
},t^{\prime }-t)\Big\}\,,
\end{equation}
where
%
%
\begin{eqnarray}
\bar{\varrho}(t_{1},t_{2}) &=& \exp \Big\{-\phi (t_{1})-\int
d^{3}rd^{3}p\,F_{1}({\mathbf{r}},{\mathbf{p}};t_{1}) \times
\notag \\
&& \hat{n}_{1}({\mathbf{r}},
{\mathbf{p}{|t}_{2})}  \notag \\
&& - \int d^{3}rd^{3}p\int d^{3}r^{\prime }d^{3}p^{\prime
}\,F_{2}({\mathbf{r}} ,{\mathbf{p}},{\mathbf{r}}^{\prime
},{\mathbf{p}}^{\prime};t_{1}) \times \notag \\
&& \hat{n} _{2}({\mathbf{r}},{\mathbf{p}},{\mathbf{r}}^{\prime
},{\mathbf{p}}^{\prime}{|t}_{2})\Big\} \, .
\end{eqnarray}
and, we recall, $\varepsilon$ is a positive real number that goes to
zero after the trace operator in the calculation of averages has
been performed [8,9], $t_{1}$ refers to the evolution in time of the
non-equilibrium thermodynamic variables and $t_{2}$ to the evolution
in time of the mechanical quantities.

The knowledge of the two distribution functions $f_{1}$ and $f_{2}$
allows to determine the value and evolution of any observable of the
system as well as of response functions and transport coefficients.
In continuation we consider the case of a unique kind of particle
(the solute) in a fluid (the solvent). The former is subjected to
external forces driving it out of equilibrium, and the latter (the
thermal bath) is taken in a steady state of constant equilibrium
with an external reservoir at temperature $T_{0}$. An analogous
case, but at the quantum mechanical level, is the one of the
carriers embedded in the ionic lattice in doped or photo-injected
semiconductors (see for example Refs. [31-34]).

We write for the Hamiltonian
%
\begin{equation}
\hat{H} = \hat{H}_{0} + \hat{H}^{\prime} \, ,
\end{equation}
where $\hat{H}_{0}$ is the kinetic energy operator and
%
\begin{equation}
\hat{H}^{\prime} = \hat{H}_{1} + \hat{W} + \hat{H}_{P} \, ,
\end{equation}
contains the internal interactions energy operator $\hat{H}_{1}$,
while $\hat{W}$ accounts for the interaction of the system with the
thermal bath, and $\hat{H}_{P}$ is the energy operator associated to
the coupling of the system with external pumping sources. They are
given by
%
\begin{equation}
\hat{H}_{0} = \int d^{3}r\int d^{3}p\frac{p^{2}}{2m}
\hat{n}_{1}({\mathbf{r}},{\mathbf{p})} + \sum_{\mu =1}^{N_{B}}
\frac{P_{\mu }^{2}}{2M} \, ,
\end{equation}
%
\begin{eqnarray}
\hat{H}_{1} &=& \frac{1}{2} \int d^{3}rd^{3}p\int d^{3}r^{\prime
}d^{3}p^{\prime }V(\left\vert \mathbf{r}-\mathbf{r}^{\prime
}\right\vert) \hat{n}_{2}({
\mathbf{r}},{\mathbf{p}},{\mathbf{r}}^{\prime},{\mathbf{p}}^{\prime
}) \notag \\
&& + \frac{1}{2} \sum_{\mu \neq \nu =1}^{N_{B}}\Phi (\left\vert
\mathbf{R}_{\mu} - \mathbf{R}_{\nu }\right\vert ) \, ,
\end{eqnarray}
%
\begin{equation}
\hat{W} = \int d^{3}r\int d^{3}p\sum_{\mu =1}^{N_{B}} w(\left\vert
\mathbf{r} - \mathbf{R}_{\mu}\right\vert
)\hat{n}_{1}({\mathbf{r}},{\mathbf{p})} \, ,
\end{equation}
%
\begin{equation}
\hat{H}_{P} = \int d^{3}r\int d^{3}p \, U_{P}(\mathbf{r},t)
\hat{n}_{1}({\mathbf{r}} ,{\mathbf{p})} \, ,
\end{equation}
where $V$ and $\Phi$ stand for the interaction potential between
two-particles in the system and in the thermal bath, respectively;
$w$ is the interaction potential between system and thermal bath,
and $U_{P}$ is the potential describing the interaction of the
system with the external pumping source; $\mathbf{R}_{\mu}$ and
$\mathbf{P}_{\mu}$ are the position and linear momentum of the
particle of mass $M$ in the thermal bath and $m$ is the mass of the
particles in the fluid.

But, for simplicity, considering a dilute solution (large distance
in average between the particles) or that the potential $V$ is
screened (e.g., molecules in a saline solvent, e.g. [35]), we can
disregard $\widehat{n}_{2}$, retaining only $\widehat{n}_{1}$. In
that case, we choose the single particle density
$\widehat{n}_{1}(\mathbf{r},\mathbf{p}|\Gamma)$ as the only relevant
dynamical variable required. Hence, the auxiliary non-equilibrium
statistical operator for the particles embedded in the bath is
\begin{equation*}
\overline{\varrho}(t,0) = \exp \Big\{ - \phi (t) - \int d^{3}r\int
d^{3}p F_{1}( \mathbf{r},\mathbf{p};t)
\widehat{n}_{1}(\mathbf{r},\mathbf{p}) \Big\} \, ,
\end{equation*}
%
\begin{equation}
\overline{\varrho}(t,0) = \prod\limits_{j=1}^{N}
\overline{\varrho}_{1}(\mathbf{r}_{j},\mathbf{p}_{j};t,0) \, ,
\end{equation}
with
%
\begin{eqnarray}
\overline{\varrho}_{1}(\mathbf{r}_{j},\mathbf{p}_{j};t,0) &=& \exp
\Big\{ - \phi_{j}(t) - \int d^{3}r \int d^{3}p
\times \notag \\
&& F_{1}(\mathbf{r},\mathbf{p};t) \delta (\mathbf{r} -
\mathbf{r}_{j}) \delta (\mathbf{p} - \mathbf{p}_{j}) \Big\}
\end{eqnarray}
which is a probability distribution in phase space for an individual
particle, with $\phi(t)$ and $\phi_{j}(t)$ ensuring the
normalization conditions of $\overline{\varrho}$ and
$\overline{\varrho}_{1}$.

The HOGH built in the framework of NESEF-kinetic theory is
characterized by the microdynamical variables consisting of the
density particles $\widehat{n} (\mathbf{r})$, the density of energy,
$\widehat{h}(\mathbf{r})$, and their fluxes of all order, i.e., the
tensors of rank $\ell $, $\widehat{I} _{n}^{\,[\ell]}(\mathbf{r})$,
and $\widehat{I}_{h}^{\,[\ell]}(\mathbf{r})$, with $\ell =1,2,...$,
given by
%
\begin{equation}
\widehat{n}(\mathbf{r}) = \int d^{3}p \,
\widehat{n}_{1}(\mathbf{r},\mathbf{p}) \, ,
\end{equation}
%
\begin{equation}
\widehat{h}(\mathbf{r}) = \int d^{3}p \frac{p^{2}}{2m}
\widehat{n}_{1}(\mathbf{r} ,\mathbf{p}) \, ,
\end{equation}
%
\begin{equation}
\widehat{\mathbf{I}}_{n}(\mathbf{r}) = \int d^{3}p \,
\frac{\mathbf{p}}{m} \widehat{n}_{1}(\mathbf{r},\mathbf{p}) \, ,
\end{equation}
%
\begin{equation}
\widehat{\mathbf{I}}_{h}(\mathbf{r}) = \int d^{3}p \,
\frac{p^{2}}{2m}\frac{
\mathbf{p}}{m}\widehat{n}_{1}(\mathbf{r},\mathbf{p}) \, ,
\end{equation}
%
\begin{equation}
\widehat{I}_{n}^{[\ell]}(\mathbf{r}) = \int d^{3}p \,
\mathbf{u}^{[\ell]}(
\mathbf{p})\widehat{n}_{1}(\mathbf{r},\mathbf{p}) \, ,
\end{equation}
%
\begin{equation}
\widehat{I}_{h}^{[\ell]}(\mathbf{r}) = \int d^{3}p \frac{p^{2}}{2m}
\, \mathbf{u}^{[\ell
]}(\mathbf{p})\widehat{n}_{1}(\mathbf{r},\mathbf{p}) \, ,
\end{equation}
where
%
\begin{equation}
\mathbf{u}^{[l]}(\mathbf{p}) = \left[
\frac{\mathbf{p}}{m}...(l-\mathrm{times}
)...\frac{\mathbf{p}}{m}\right] \, .
\end{equation}
is the tensorial product of $\ell$ vectors $\mathbf{p}/m$, defining
a $\ell$-\emph{rank tensor} (it has the dimensions of velocity to
the power $\ell$).

The average over the non-equilibrium ensemble of the quantities
above
provides the set of macrovariables,
%
\begin{equation}
\left\{ n(\mathbf{r},t), \, \mathbf{I}_{n}(\mathbf{r},t), \,
\{I_{n}^{[\ell]}( \mathbf{r},t)\} \right\} \, ,
\end{equation}
which we call the \emph{family of variables describing the particle motion},
and
%
\begin{equation}
\left\{ h(\mathbf{r},t), \, \mathbf{I}_{h}(\mathbf{r},t), \,
\{I_{h}^{[\ell]}( \mathbf{r},t)\} \right\} \, ,
\end{equation}
which we call the \emph{family of variables describing the heat
motion}, which define the HOGH.

It can be noticed that in the present case those associated to the
energy are contained in those associated to the movement of matter.
Beginning with $I_{n}^{[2]}$ the contraction of the two first
indexes provide the expression for the energy, such contraction in
$I_{n}^{[3]}$ results in the flux of energy (heat current), and so
on.

According to Eqs. (3) and (13) we have that
%
\begin{equation}
f_{1}(\mathbf{r},\mathbf{p};t) = \mathrm{Tr} \left\{
\widehat{n}_{1}(\mathbf{r}, \mathbf{p}) \prod\limits_{j=1}^{N}
\overline{\varrho}_{1}(\mathbf{r}_{j}, \mathbf{p}_{j};t,0) \right\}
\, ,
\end{equation}
is NESEF single-particle distribution function is space
$(\mathbf{r},\mathbf{p})$; $\mathrm{Tr}$ indicates integration in
phase space and we recall that the average values of the basic
dynamical values, and only for them, there is coincidence of the one
calculated with the statistical operator
$\varrho_{\varepsilon}(t,0)$ and the calculation with the auxiliary
``instantaneous quasi-equilibrium distribution" $\overline{\varrho
}(t,0)$.

Hence, the basic macrovariables of HOGH, that is, the average values
of the microdynamical quantities in Eqs. (15) to (20) are expressed
in terms of the single-particle distribution, namely

1. For the family describing particle motion
%
\begin{equation}
I_{n}^{[\ell]}(\mathbf{r},t) = \int d^{3}p\,u^{[\ell]}(\mathbf{p})
f_{1}(\mathbf{r},\mathbf{p};t) \, ,
\end{equation}

2. For the family describing heat motion
%
\begin{equation}
I_{h}^{[\ell]}(\mathbf{r},t) = \int
d^{3}p\,\frac{p^{2}}{2m}u^{[\ell]}( \mathbf{p})
f_{1}(\mathbf{r},\mathbf{p};t) \, ,
\end{equation}
with $\ell = 0$ corresponding to the densities, $n(\mathbf{r},t)$
and $h(\mathbf{r},t)$, $\ell = 1$ to the first (vectorial) fluxes,
$\ell \geqslant 2$ to the higher order fluxes. Moreover, it can be
shown that we can write for the flux
%
\begin{equation}
\mathbf{I}_{n}(\mathbf{r},t) =
n(\mathbf{r},t)\mathbf{v}(\mathbf{r},t) \, ,
\end{equation}
where $\mathbf{v}(\mathbf{r},t)$ is the field of barycentric velocity, and
for the second-order flux that
%
\begin{equation}
mI_{n}^{[2]}(\mathbf{r},t) = P^{[2]}(\mathbf{r},t) +
m[\mathbf{v}(\mathbf{r},t) \mathbf{v}(\mathbf{r},t)] \, ,
\end{equation}
where $P^{[2]}$ is the pressure tensor and the last term on the
right is the convective pressure tensor. Next we proceed to derive
in the HOGH context the kinetic equations that govern the
hydrodynamic motion.


\section{Evolution Equations in NESEF-HOGH}

First we noticed that in NESEF-based kinetic theory [8,12,36,37] the
evolution equation for a macrovariable $A(\mathbf{r},\mathbf{p},t)$,
being the average over the non-equilibrium ensemble of the
microdynamical variable $\hat{A}(\mathbf{r},\mathbf{p})$, consists
into the average over such ensemble of the mechanical equation of
motion, namely,
%
\begin{eqnarray}
\frac{\partial}{\partial t}A(\mathbf{r},\mathbf{p},t) &=&
\frac{\partial}{
\partial t} \mathrm{Tr} \{ \hat{A}(\mathbf{r},\mathbf{p}) \varrho_{\varepsilon
}(t)\} = \notag \\ &=& \mathrm{Tr} \{
\{\hat{A}(\mathbf{r},\mathbf{p}),\hat{H}\} \varrho_{\varepsilon}(t)
\} \, ,
\end{eqnarray}
where $\{...,...\}$, stands for Poisson bracket. A practical way to
handle this equation, in the form of a perturbative series, ordered
according to increasing powers of the interaction strength, is fully
described in Refs. [8,37]. The successive terms in the series invoke
pair collisions, triple and higher-order collisions, all of them
including memory and vertex renormalization. A truncation of the
series in lowest order (keeping only the terms quadratic in the
interaction strengths) is a Markovian approximation [37,38].

From now on we concentrate the attention on the evolution of the thermal
quantities
%
\begin{equation}
\left\{ h(\mathbf{r},t), \, \mathbf{I}_{h}(\mathbf{r},t), \,
\{I_{h}^{[\ell]}(\mathbf{r},t)\} \right\} \, ,
\end{equation}
with $\ell =2,3,\ldots $, that is, the density of energy and its fluxes of
all orders given in Eq. (26). The evolution equations are:
%
\begin{equation}
\frac{\partial}{\partial t} I_{h}^{[\ell]}(\mathbf{r},t) = \int
d^{3}p\,\frac{ p^{2}}{2m}u^{[\ell]}(\mathbf{p}) \frac{\partial
}{\partial t} f_{1}(\mathbf{r} ,\mathbf{p};t) \, ,
\end{equation}
in which we need to introduce the kinetic equation for the
single-particle $f_{1}(\mathbf{r},\mathbf{p};t)$ which is given in
Ref. [25] and briefly described in Appendix A. Performing the
lengthy calculations involved we finally arrive at the general
evolution equations ($\ell = 0,1,2,\ldots $)
%
\begin{equation}
\frac{\partial}{\partial t}I_{h}^{[\ell]}(\mathbf{r},t) + \nabla
\cdot I_{h}^{[\ell +1]}(\mathbf{r},t) = -
\boldsymbol{\mathcal{F}}(\mathbf{r} ,t) I_{n}^{[\ell
+1]}(\mathbf{r},t) +  \notag
\end{equation}
\begin{equation}
-\sum_{s=1}^{\ell}\wp (1,s) \left[
\boldsymbol{\mathcal{F}}(\mathbf{r},t) I_{h}^{[\ell
-1]}(\mathbf{r},t) \right] + \theta_{\ell}^{-1}I_{h}^{[\ell
]}(\mathbf{r},t) +  \notag
\end{equation}
\begin{equation*}
+a_{L0} \sum_{s=1}^{\ell} \wp (1,s) \left[ \nabla I_{n}^{[\ell
-1]}(\mathbf{r},t) \right] +
\end{equation*}
\begin{equation}
+2(\ell + 3) a_{L1} \nabla \cdot I_{h}^{[\ell +1]}(\mathbf{r},t) +
S_{h}^{[\ell]}(\mathbf{r},t) \, ,
\end{equation}
where
%
\begin{equation}
\theta_{\ell}^{-1} = (\ell +2) \left\vert a_{\tau 0} \right\vert + [
\ell^{2} + 5(\ell +1) \left\vert b_{\tau 1} \right\vert ] \, ,
\end{equation}
is the reciprocal of the Maxwell time [39,40] associated to the
$\ell$-th order flux. Moreover
%
\begin{eqnarray}
S_{h}^{[\ell]}(\mathbf{r},t) &=& (2\ell +3) b_{\tau _{0}}\frac{m}{2}
I_{h}^{[\ell]}(\mathbf{r},t) + \notag \\
&& b_{\tau_{0}} \left\{ \widehat{\wp }_{\ell} \left[
\mathbf{1}^{[2]} I_{h}^{[\ell -2]}(\mathbf{r},t) \right] \right\} +
\notag \\
&& b_{\tau_{1}} \frac{2}{m} \left\{ \widehat{\wp }_{\ell} \left[
\mathbf{1}^{[2]} I_{h_{2}}^{[\ell - 2]}(\mathbf{r},t) \right]
\right\} + \notag \\
&& a_{L_{0}} m \nabla \cdot I_{n}^{[\ell + 1]}(\mathbf{r},t) + \notag \\
&& + a_{L_{1}} \frac{2}{m} \sum_{s=1}^{\ell} \wp (1,s) \left[ \nabla
I_{h_{2}}^{[\ell - 1]}(\mathbf{r},t) \right] + \notag \\
&& 3(\ell + 2) a_{\tau_{1}} \frac{2}{m
}I_{h_{2}}^{[\ell]}(\mathbf{r},t) + R_{h}^{[\ell]}(\mathbf{r},t) \,
,
\end{eqnarray}
is a contribution which couples the thermal motion to the material
motion (thermal-striction effect), $R_{h}^{[\ell]}(\mathbf{r},t)$
contains the contribution of all the other higher-order fluxes
($\geqslant \ell +2$). The several kinetic coefficients are
presented in Appendix B. Taking into account that (see Appendix B)
%
\begin{equation}
|b_{\tau_{1}}| = \frac{1}{5 (1+x)} |a_{\tau_{0}}| \, ,
\end{equation}
it follows that Maxwell times have the property that
%
\begin{equation}
\frac{\theta_{\ell +1}}{\theta_{\ell}} = \frac{5(\ell +2) \left( 1 +
x \right) + \ell^{2} + 5(\ell +1)}{5(\ell +3)(1+x) + (\ell
+1)^{2}+5(\ell + 2)} < 1 \, ,
\end{equation}
for $\ell =0,1,2,3,...$ and where $x=m/M$. The ordered sequence
%
\begin{equation}
\theta_{0} > \theta_{1} > \theta_{2} > \theta_{3} > ... >
\theta_{\ell} > \theta_{\ell + 1} > ...
\end{equation}
is verified, and may be noticed that for large $\ell$,
$\theta_{\ell}$ goes as $\ell^{-2}$, i.e., going to zero as $\ell
\rightarrow \infty$.

For the Brownian particle ($m/M \gg 1$) the ratio in Eq. (36) tends
asymptotically to $(\ell + 2)/(\ell + 3)$, while for Lorentz
particles ($m/M \ll 1$) the ratio tends to $[\ell^{2} + 5(2\ell +
3)/[(\ell + 1)^{2} + 5(2\ell + 5)]$. Considering that the Maxwell
time of the first flux of matter, $\theta_{n_{1}}$ (which is the
linear momentum relaxation time), in Ref. [24], is
$|a_{\tau_{0}}|^{-1}$, we can establish the relation between it and
the Maxwell time for the energy, namely
%
\begin{equation}
\theta_{0} = \frac{\left( 1+x\right)}{2x+3} \theta_{n_{1}} \, ,
\end{equation}
what tells us that $\theta_{0} < \theta_{n_{1}}$. For Brownian
particles $\theta_{0} = \theta_{n_{1}}/2$ and for the Lorentz
particles $\theta_{0} = \theta_{n_{1}}/3$. Moreover, for general
$\ell$ we have that
%
\begin{equation}
\theta_{\ell} = \frac{5 \left(1 + x \right)}{5(\ell + 2)(1 + x) +
\ell^{2} + 5(\ell + 1)} \theta_{n_{1}} \, .
\end{equation}
%


\section{Contraction of Description}

Equation (32) represents the coupled set of evolution equations
involving the density and all its fluxes in its most general form.
It must be noticed that it is linear in the basic variables; no
approximation has been introduced. Nonlinearities should arise out
of the inclusion of inter-particle interaction which we have
disregarded in the present communication (case of a dilute
solution). However, as already noticed, such set of equations is
intractable, and, of course, we need to look in each case on how to
find the best description using the smallest possible number of
variables. In other words to introduce an appropriate -- for each
case -- contraction of description: this contraction implies in
retaining the information considered as relevant for the problem in
hands, and to disregard non-relevant information [41].

Elsewhere [42] we have discussed the question of the contraction of
description (reduction of the dimensions of the non-equilibrium
thermodynamic space of states). As shown a criterion for justifying
the different levels of contraction is derived: It depends on the
range of wavelengths and frequencies which are relevant for the
characterization, in terms of normal modes, of the
hydro-thermodynamic motion in the non-equilibrium open system.
Maxwell times have a particular relevance for establishing such
contractions, for which a truncation criterion can be derived, which
rests on the characteristics of the hydrodynamic motion that
develops under the given experimental procedure.

Since inclusion of higher and higher order fluxes implies in
describing a motion involving increasing Knudsen numbers per
hydrodynamic mode (that is governed by smaller and smaller
wavelengths -- larger and larger wave numbers -- accompanied by
higher and higher frequencies), in a qualitative manner, we can say,
as a general ``thumb rule", that the criterion indicates that
\emph{a more and more restricted contraction can be used when larger
and larger are the prevalent wavelengths in the motion}. Therefore,
in simpler words, when the motion becomes more and more smooth in
space and time, the more reduced can be the dimension of the basic
macrovariables space to be used for the description of the
non-equilibrium thermodynamic state of the system.

As shown elsewhere [42], it can be conjectured a general criterion
of contraction, namely, a truncation of order $r$ (meaning keeping
the densities and their fluxes up to order $r$) can be introduced,
once we can show that in the spectrum of wavelengths, which
characterizes the motion, predominate those larger than a
``frontier" one, $\lambda_{(r,r+1)}^{2} = v^{2} \theta_{r}
\theta_{r+1}$, where $v$ is of the order of the thermal velocity and
$\theta_{r}$ and $\theta_{r+1}$ the corresponding Maxwell times
associated to the fluxes of order $r$ and $r + 1$. We shall try to
illustrate the matter using a contraction of order 1.


\section{HOGH of Order 1}

In this section, we consider very briefly a relation of the theory
so far presented and the analysis of a particular techno-industrial
process, namely, laser thermal stereolithography. Stereolithography
is a technological process that allows solid parts to be made
directly and rapidly from computer data (prototyping) without the
necessity of tooling or cutting machining. The presently available
industrial process is based on the use of photopolymers, with the
solid sections built using laser ultraviolet light which induces a
photochemical reaction solidifying the resin and thus providing a
prototype (matrix) of the part of the object (in machines, cars,
medical implants, etc.) to be produced. An alternative procedure
involves the use of thermo sensitive resins, like some waxes and
compounds, and with them the solid sections are formed by a
sintering process produced by application of heat generated by an
infrared-light laser [43]. This alternative rapid prototyping method
has the economic advantage of using much cheaper and easily
available thermo-sensitive resins instead of the rare and expensive
photosensitive resins used with the other, original, approach. We
consider next some hydrodynamic aspects of this technique.

It is essential to the process that the produced protypes be
homogeneous and with well-defined and satisfactorily precise
geometry, leading to spatially well-defined solid parts with
three-dimensional geometry. Advances in such direction are being
obtained, which are related to quite interesting and illustrative
aspects of the generalized hydrodynamics so far described.

In a typical experiment, a beam originating from a CO$_{2}$ laser is
used ($\lambda $ = 10.6 $\mu$m with power, for best technical
results, around 25 watts). An optical system directs the laser beam
to illuminate and heat the polymeric resin with the laser beam
focused to a spot of diameter, say, $2r_{0}$ (usually in the 1 mm
range). The resin (e.g., epoxy, polyester) together with a curing
agent (e.g., dimetiltriamine), compounding the media where the
sintering process is to proceed, is then locally heated and a
solidification follows. The optical system governed by the computer
program focuses the beam along the correct path for the prototype to
be produced. However, the experience in the laboratory showed that
using a pure material (the resin and the curing substance) the
process was not rapid or well localized. The pressure and heating
originated on the spot of focalization of the laser beam generated a
material wave accompanied by heat motion, resulting in the
solidification (sintering) of the resin over large extensions and
thus ruining the purpose of the technique that, as noted, requires a
quite localized solidification to obtain a well-shaped model. This
drawback was corrected to a large degree by resorting to mix the
resin with silica (or aluminum) powder with the grains acting as
collision centers.

The theory presented in the previous section allows us to analyze
and understand those results. In the process are present, evidently,
matter and heat motion in the body of the sample, but neglecting
thermostriction effects (which seems a plausible approximation in
the case under consideration), we can apply the theory just
presented. Hence, material and thermal motion are decoupled and we
concentrate our attention on the equations of motion for energy,
whose propagation is of fundamental relevance in this technological
process.

Let us consider in the general equation (32) the first two ones,
$\ell = 0$ and $\ell = 1$, in the approximate forms, disregarding
$\boldsymbol{\mathcal{F}}(\mathbf{r},t)$,
%
\begin{equation}
\frac{\partial}{\partial t} h(\mathbf{r},t) + \nabla \cdot
\mathbf{I}_{h}(\mathbf{r},t) = - \frac{1}{\theta_{0}}
h(\mathbf{r},t) \, ,
\end{equation}
%
\begin{eqnarray}
\frac{\partial}{\partial t} \mathbf{I}_{h}(\mathbf{r},t) + \nabla
\cdot I_{h}^{[2]}(\mathbf{r},t) &=& - \frac{1}{\theta
_{1}}\mathbf{I}_{h}(\mathbf{r},t) + a_{L0} \nabla h(\mathbf{r},t) +
\notag \\
&& 8 a_{L1} \nabla \cdot I_{h}^{[2]}(\mathbf{r},t) \, ,
\end{eqnarray}
in the interparticle interaction and the presence of impurities.

Differentiating in time Eq. (40) and next using Eq. (41), after
ignoring the therm $\nabla \cdot \nabla \cdot I_{h}^{[2]}$ we arrive
to a closed equation for the propagation of energy, namely,
%
\begin{equation}
\left( \frac{1}{c_{\varepsilon}^{2}} \frac{\partial ^{2}}{\partial
t^{2}} + \frac{1}{D_{\varepsilon}} \frac{\partial}{\partial t} -
\nabla ^{2}\right) h( \mathbf{r},t) = -
\frac{h(\mathbf{r},t)}{\varkappa_{\varepsilon} \theta_{1}} \, ,
\end{equation}
where
%
\begin{equation}
c_{\varepsilon}^{2} = \frac{\varkappa_{\varepsilon}}{\theta _{1}} \,
,
\end{equation}
is the velocity of propagation, and
%
\begin{equation}
D_{\varepsilon} = \frac{\varkappa_{\varepsilon}
\bar{\theta}}{\theta_{1}} \, ,
\end{equation}
is a diffusion coefficient where $\bar{\theta}^{-1} =
\theta_{0}^{-1} + \theta_{1}^{-1}$ and $\varkappa_{\varepsilon} =
\theta_{1} (1- 6a_{L1})a_{L0}$. Except for the relaxation term on
the right, Eq. (42) is an hyperbolic equation of the telegraphist
type, referred to as Maxwell-Cattaneo equation. It implies in the
propagation of a small amplitude motion of heat on the background
provided by the original state of thermal equilibrium at room
temperature.

Neglecting the right side of Eq. (42) (i.e., the term that leads to
the cooling of the system after a certain time interval has
elapsed), we solve it for the case of particular initial and
boundary conditions. They consist of taking a laser pulse producing
a concentration of energy at the spot of focalization, which is
distributed on the material upper thin layer with a
Gaussian profile of width $r_{0} \simeq $ 1.0 mm, i.e.,
%
\begin{equation}
h(r,0) = A(\pi /r_{0}^{2}) e^{-r^{2}/4r_{0}^{2}} \, ,
\end{equation}
where $A$ is an amplitude related to the intensity of the pumped
laser light energy. In Eq. (45), $r$ is the planar radius and we
assume a near uniform heating in depth (perpendicular direction
$z$). Hence, we can separate the planar polar coordinates from the
$z$-coordinate and use the initial and boundary condition of Eq.
(45) for solving the resulting equation, whose solution we write in
the form of a superposition of normal modes on the surface sample,
namely:
%
\begin{equation}
h(\mathbf{r},t) = \int d^{2}k \, \tilde{\varepsilon}(\mathbf{k}) e^{
i(\mathbf{k} \cdot \mathbf{r} - \omega_{k}t)} \, ,
\end{equation}

Inserting Eq. (46) in Eq. (42), we obtain the characteristic
equation that defines the energy dispersion relation in the
propagation of this second sound, which is
%
\begin{equation}
\omega_{k} = - \frac{i}{\tau_{\varepsilon }}\pm
\sqrt{c_{\varepsilon}^{2} k^{2} - 1/(\tau_{\varepsilon})^{2}} \, ,
\end{equation}
where $\tau_{\varepsilon} = 2 D_{\varepsilon}/c_{\varepsilon}^{2}$
is the lifetime.

We can see that depending on the value of $k$ this dispersion relation
separates two types of regimes, namely:
%
\begin{equation}
k > \frac{1}{ c_{\varepsilon} \tau_{\varepsilon} } \equiv k_{c} \, ,
\end{equation}
which corresponds to a damped (with lifetime $\tau_{\varepsilon}$)
undulatory motion, while for
%
\begin{equation}
k < \frac{1}{ c_{\varepsilon} \tau_{\varepsilon}} \equiv k_{c} \, ,
\end{equation}
there follows an overdamped movement, which, as discussed below, consists of
a diffusive motion. Moreover in the case $k < k_{c}$, approximately
%
\begin{equation}
\omega_{k} = \Bigg\{
\begin{array}{c}
- i D_{\varepsilon} k^{2} \\
- i/\tau_{\varepsilon}    \\
\end{array}
\end{equation}
while for $ k > k_{c}$, we can write that
%
\begin{equation}
\omega_{k} = - i/\tau_{\varepsilon} \pm \tilde{\omega}_{k} \, ;
\qquad \tilde{\omega}_{k} = c_{\varepsilon} \sqrt{k^{2} - k_{c}^{2}}
\, ,
\end{equation}
with $\tilde{\omega}$ being a real number frequency. Hence, on the
basis of these results, we can separate Eq. (46) in the form
%
\begin{eqnarray}
h(\mathbf{r},t) &=& \int_{k<k_{c}} d^{2} k[a(\mathbf{k})
e^{-t/\tau_{\varepsilon}} + \tilde{h}(\mathbf{k}) e^{-D_{\varepsilon
}k^{2}t}] e^{i
\mathbf{k} \cdot \mathbf{r}} +  \notag \\
&& + \int_{k>k_{c}} d^{2}k \tilde{h}(\mathbf{k})
e^{-t/\tau_{\varepsilon}} \cos (\tilde{\omega}_{k}t) e^{i\mathbf{k}
\cdot \mathbf{r}} \, ,
\end{eqnarray}
where $a(\mathbf{k})$ and $\tilde{h}(\mathbf{k})$ are coefficients
to be determined by the initial and boundary conditions of Eq. (45).
The latter can be Fourier analyzed to obtain that
%
\begin{equation}
h(\mathbf{r},0) = A(\pi /r_{0}^{2}) \int d^{2} k \,
e^{-k^{2}r_{0}^{2}} e^{-\mathbf{k} \cdot \mathbf{r}} \, ,
\end{equation}
and then
%
\begin{equation}
\tilde{h}(\mathbf{k}) = A(\pi /r_{0}^{2}) e^{-k^{2}r_{0}^{2}} \, ,
\end{equation}
and we are neglecting in Eq. (52) the term with coefficient
$a(\mathbf{k})$ because of the rapid decay of this term in
comparison with the other in this region where $k < k_{c}$.
Therefore, we obtain that
%
\begin{eqnarray}
h(\mathbf{r},t) &=& A(\pi /r_{0}^{2}) \Big[ \int_{k<k_{c}} d^{2}k \,
e^{-k^{2}r_{0}^{2}} e^{-D_{\varepsilon
}k^{2}t} e^{i\mathbf{k} \cdot \mathbf{r}} +  \notag \\
&& + \int_{k>k_{c}} d^{2}k \, e^{-k^{2}r_{0}^{2}} e^{-t/\tau
_{\varepsilon }} \cos ( \tilde{\omega}_{k}t) e^{i\mathbf{k} \cdot
\mathbf{r}} \Big] \, .
\end{eqnarray}

Equation (55) clearly tells us that the movement, as already
noticed, is composed of a diffusive motion associated to the modes
with long wavelengths ($\lambda > c_{\varepsilon}
\tau_{\varepsilon}$) and damped undulatory motion associated to the
intermediate to short wavelengths ($\lambda < c_{\varepsilon}
\tau_{\varepsilon}$), similarly, to the case of the telegraphist
equation in electrodynamics.

Inspection of Eq. (55) leads us to the conclusion that, on account
of the Gaussian envelope in the integral, the contributions to the
integrals for values of $k$ such that $k r_{0} > 1$ are negligible.
Therefore, considering the limit that separates the undulatory and
diffusive regimes namely, $k_{c} = (c_{\varepsilon} \tau
_{\varepsilon})^{-1}$, which we write $k_{c} \lambda_{\varepsilon} =
1$; where $\lambda_{\varepsilon} = c_{\varepsilon}
\tau_{\varepsilon}$ is the mean free path, we can notice that:

(i) If $k_{c}r_{0} = r_{0}/\lambda_{\varepsilon} < 1$, the motion is
composed of a diffusive one accompanied by a wave that carries heat
at speed $c_{\varepsilon}$; and similarly for the expected movement
of mass (a material or pressure wave) that expands from the initial
burning hole produced by the incidence of the laser pulse (see Fig.
1), and

(ii) if $k_{c} r_{0} = r_{0}/\lambda_{\varepsilon} > 1$, then the
movement is purely diffusive, and in this case,
%
\begin{equation}
h(\mathbf{r},t) = \frac{\pi A}{r_{0}^{2} + D_{\varepsilon }t} \,
e^{-r^{2}/(r_{0}^{2} + D_{\varepsilon }t)} \, ,
\end{equation}
which is illustrated in Figure 2.

\begin{figure}[h]
\center
\includegraphics[angle=180,width=8.5cm]{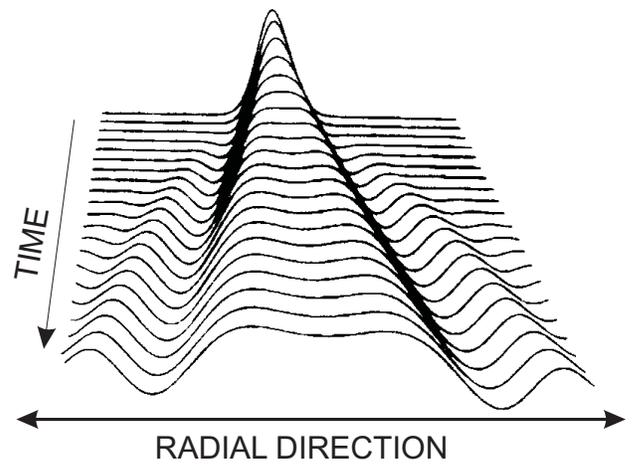}
\caption{Description of the domain of propagation of a diffusive
character accompanied with undulatory motion. The space dependence
amplitude of the density for increasing time indicated by the
arrow.}
\end{figure}

\begin{figure}[h]
\center
\includegraphics[width=8.5cm]{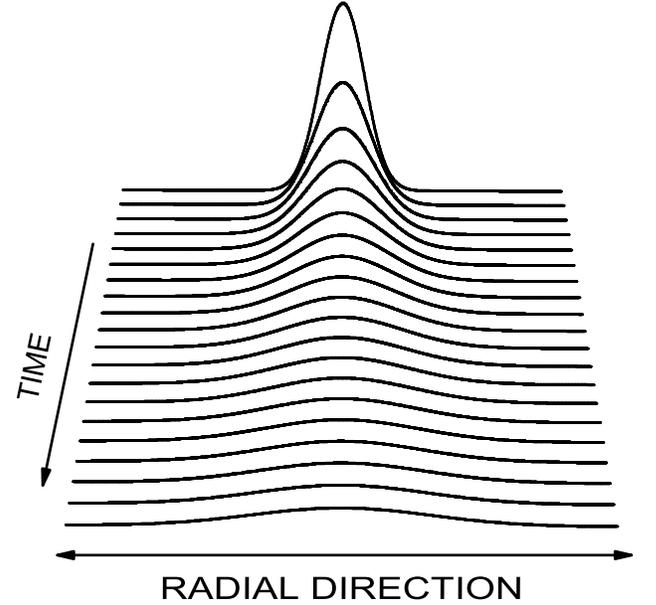}
\caption{Description of the domain of propagation of a pure
diffusive motion. The space dependence amplitude of the density for
increasing time indicated by the arrow.}
\end{figure}

On the basis of the analysis above, the prototype in the case when
the pure resin is used is ruined, as noticed, because of
solidification over a large area, i.e., a large area around the spot
of laser beam focalization has been heated. We conclude that, in
this case, the situation should be such that $k_{c} r_{0} < 1$ and
the undulatory motion, before decaying, carries heat at distances,
in the experiments, of the order of centimeters. On the other hand,
addition of silica powder greatly improves the situation, producing
an acceptable prototype. We conclude that when $k_{c} r_{0} > 1$ the
wave propagation of heat is not present, only slow diffusion
accompanied by rapid cooling and sinterization on a small space
scale (millimeter in the experiment).

That the presence of silica powder increases $k_{c}$ can be
understood on the basis that $k_{c} = (c_{\varepsilon}
D_{\varepsilon})^{-1} = \lambda_{\varepsilon}^{-1}$, and the
presence of the macroscopic grains of silica decreases the mean free
path $\lambda_{\varepsilon}$, increasing therefore $k_{c}$. It
should be noticed that also the diffusivity $D_{\varepsilon}$
decreases, further improving the results.

Consequently, we can say that the theoretical analysis we performed
appears as justified, which corroborates the observed experimental
behavior. The study of this industrial and commercial technique, on
the basis of the simplest form of the approach to the generalized
hydrodynamics described here (i.e., only one step further from the
domain of classical hydrodynamics), allows one, on the one hand, to
discuss and reinforce some aspects of the theory, and, on the other,
to visualize the physical phenomena developing in the process and
then to analyze in depth the problem and look for optimization of
the technique.


\section{Concluding Remarks}

In summary, the study in depth of the characteristics and responses of
materials working, as in devices and industrial processes, under
far-from-equilibrium conditions and, eventually, involving complex and
nanometric conditions, can presently be performed in terms of two main
formalisms, namely, computational modelling [4,5] or a generalized kinetic
theory. We have here presented the derivation, based on a Gibbs-style
nonequilibrium statistical ensemble formalism, of a kinetic theory which
permitted the construction of a generalized kinetic equation having an ample
domain of validity and large number of possible applications.

In this communication we have worked at the classical mechanical
level. The quantum mechanical approach follows along quite similar
lines; in that case the classical single-particle distribution
$f_{1}(\mathbf{r},\mathbf{p};t)$ is replaced by the Wigner-von
Neumann one and, clearly, Heisenberg quantum equations of motion
enter in place of the Hamiltonian classical ones. The quantum
approaches has been described and applied in the case of
semiconductors, where it has an important paper in the determination
of heat conductivity at the nanometric scale, when it is evidenced
an important deviation from the value that follow in the standard
approach [44-47]. The so called Maxwell characteristic times [39,48]
were evidenced and their expressions derived. They are of a
fundamental relevance to characterize the type of motion and,
therefore, to determine the choice of the contracted description to
be used in practical applications as described in Section IV.

\appendix{}

\section{NESEF-Kinetic Equation}

A kinetic equation of evolution for the single-particle distribution
function $f_{1}(\mathbf{r},\mathbf{p};t)$ follows from the
NESEF-based kinetic theory [8,12,37],
%
\begin{equation}
\frac{\partial}{\partial t}f_{1}(\mathbf{r},\mathbf{p};t) =
\mathrm{Tr} \left\{
\{\widehat{n}_{1}(\mathbf{r},\mathbf{p}),\widehat{H}\}
\varrho_{\varepsilon}(t) \right\} \, ,
\end{equation}
that is, the average over the nonequilibrium ensemble characterized
by the statistical operator $\varrho_{\varepsilon}(t)$.

Using the Hamiltonian of Eq. (7), with $\widehat{H}_{0}$ being the kinetic
energy and $\widehat{H}^{\prime }$ containing all the interactions, that is,
pair interactions between the system particles, of the system particles with
those of the thermal bath, and the interactions with applied external
sources. Resorting to the approximation of keeping only the contribution
from the irreducible two particle collisions, lengthy but straightforward
calculations lead to the kinetic equation
%
\begin{equation*}
\frac{\partial}{\partial t} f_{1}(\mathbf{r},\mathbf{p};t) +
\frac{\mathbf{P}(\mathbf{r},\mathbf{p};t)}{m} \cdot \nabla
f_{1}(\mathbf{r},\mathbf{p};t) +
\end{equation*}
\begin{equation*}
+\mathbf{F}(\mathbf{r},\mathbf{p};t) \cdot \nabla_{\mathbf{p}}
f_{1}(\mathbf{r},\mathbf{p};t) - B(\mathbf{p})
f_{1}(\mathbf{r},\mathbf{p};t) +
\end{equation*}
\begin{equation*}
-A_{2}^{[2]}(\mathbf{p}) \odot \lbrack \nabla_{\mathbf{p}} \nabla]
f_{1}(\mathbf{r},\mathbf{p};t) +
\end{equation*}
\begin{equation}
- B_{2}^{[2]}(\mathbf{p}) \odot \lbrack \nabla_{\mathbf{p}}
\nabla_{\mathbf{p}}] f_{1}(\mathbf{r},\mathbf{p};t) =
J_{S}^{(2)}(\mathbf{r},\mathbf{p};t) \, ,
\end{equation}
where
%
\begin{equation}
\mathbf{P}(\mathbf{r},\mathbf{p};t) = \mathbf{p} -
m\mathbf{A}_{1}(\mathbf{p}) \, ,
\end{equation}
plays the role of a generalized nonlinear momentum,
%
%
\begin{equation}
\mathbf{F}(\mathbf{r},\mathbf{p};t) = - \nabla U_{ext}(\mathbf{r},t)
- \mathbf{B}_{1}(\mathbf{p}) - \mathbf{F}_{NL}(\mathbf{r};t) -
\nabla U(\mathbf{r};t) \, ,
\end{equation}
is a generalized force, in which
%
%
\begin{equation}
\mathbf{F}_{NL}(\mathbf{r};t) = \int d^{3}r^{\prime} \int
d^{3}p^{\prime} \mathbf{G}_{NL}(\mathbf{r}^{\prime} -
\mathbf{r},\mathbf{p}^{\prime}) f_{1}( \mathbf{r}^{\prime
},\mathbf{p}^{\prime};t) \, ,
\end{equation}
with full details given in Ref. [24]. The symbol $\odot$ stands for
contracted product of tensors.

When Eq. (A2) is compared with, say, standard Boltzmann kinetic
equation, it contains several additional contributions. First,
$\mathbf{P}(\mathbf{p};t)$, Eq. (A3), can be interpreted as a
modified momentum, composed of the linear one, $\mathbf{p}$, plus a
contribution arising out of the interaction with the thermal bath,
$m \mathbf{A}_{1}(\mathbf{p})$. The contribution $m
\mathbf{A}_{1}(\mathbf{p})$ can be considered as implying in a
transfer of momentum between system and bath.

The force $\mathbf{F}(\mathbf{r},\mathbf{p};t)$, in Eq. (A4), is
composed of four contributions: the first one, $-\nabla U_{ext}$, is
the external applied force; next, $\mathbf{B}_{1}(\mathbf{p})$,
arising out of the interaction with the thermal bath, is a
contribution that, together with the fourth one in Eq. (A2), namely,
$\mathbf{B}(\mathbf{p}) f_{1}$, constitutes a generalization of the
so-called effective friction force. In the limit of a Brownian
system ($m\gg M$) it is recovered the known expression [49] $\gamma
\nabla_{\mathbf{p}}[(\mathbf{p}/m) f_{1}(\mathbf{r},\mathbf{p};t)]$
with the friction coefficient given by $\gamma = m/\tau$ where $\tau
$ is the momentum relaxation time. In the limit of a Lorentz system
($m\ll M$) $\mathbf{B}_{1}(\mathbf{p})$ and $\mathbf{B}(\mathbf{p})$
tend to zero and then this friction force is practically null.

The third contribution to the force
$\mathbf{F}(\mathbf{r},\mathbf{p};t)$, i.e.
$\mathbf{F}_{NL}(\mathbf{r};t)$, is an interesting one, which is an
effective force between pairs of particles generated through the
interaction of each of the pair with the thermal bath. A similar
presence of an induced effective coupling of this type has been
evidenced in the case of two Brownian particles embedded in a
thermal bath [50] and also it can be noticed the similarity with the
one that leads to the formation of Cooper pairs in type-I
superconductivity. Such contribution is of a nonlinear (bilinear in
$f_{1}$) character, and therefore eventually may lead to the
emergence of complex behavior in the system, e. g., the cases of the
so-called non-equilibrium Bose-Einstein-like condensation [51-56].
The last contribution, $- \nabla U$, is usually called the
self-consistent or mean field force, containing the proper
single-particle distribution and then providing a nonlinear
contribution to the kinetic equation.

The fifth term on the left of Eq. (A2) consists of a cross double
differentiation, namely the rank-2 tensor $[\nabla_{\mathbf{p}}
\nabla ] f_{1}$, which takes into account effects of anisotropy
caused by non-uniformity. The sixth one consists of a double
differentiation in the momentum variable, the rank-2 tensor $[\nabla
_{\mathbf{p}} \nabla_{\mathbf{p}}] f_{1}$, which is related to the
so-called diffusion in momentum space. Both contributions have their
origins in the interaction of the particles with the thermal bath;
that is in those terms of the collision integral $J_{1}^{(2)}$
containing the potential $w$ (see Eq. (11)).

Finally, on the right, $J_{S}^{(2)}$\ is the collision integral
resulting from the interaction between the particles taken in the
weak coupling limit, i.e., the so called weakly coupled gas
collision integral [57].

To go beyond the weak-coupling approximation, one needs to go back
to the general kinetic equation to include memory effects and
higher-order collision integrals, so that to include vertex
renormalization [52].


\section{The Kinetic Coefficients in Eq. (32)}

We do have that,
%
\begin{equation}
a_{\tau 0} = \frac{\mathcal{V}}{(2 \pi)^{3}} \frac{4 \pi}{3} \int dQ
\, Q^{4} f_{\tau 0}(Q) \, ,
\end{equation}
with
 %
\begin{equation}
f_{\tau 0}(Q) = - \frac{n_{R}}{\mathcal{V}} \frac{\left( M \beta_{0}
\right)^{3/2} \pi}{\sqrt{2 \pi} m^{2}} \frac{\left\vert \psi
(Q)\right\vert^{2}}{Q} \left( \frac{m}{M} + 1 \right)
\end{equation}
where $\psi (Q)$ is the Fourier transform of the potential energy $
w(\left\vert \mathbf{r}_{j}-\mathbf{R}_{\mu}\right\vert )$, $n_{R}$
is the density of particles in the thermal bath, $\mathcal{V}$ is
the volume, and $\beta_{0}^{-1} = k_{B} T_{0}$. Moreover,
%
\begin{equation}
a_{\tau 1} = - \frac{M \beta_{0}}{10} a_{\tau 0} \, ,
\end{equation}
%
\begin{equation}
a_{L0} = \sqrt{\frac{2}{M \beta_{0} \pi }} \frac{1}{\kappa} a_{\tau
0} \, ,
\end{equation}
%
\begin{equation}
\frac{1}{\kappa} = \frac{\int dQ \, Q^{2}\left\vert \psi(Q)
\right\vert ^{2} }{\int dQ \, Q^{3}\left\vert \psi(Q) \right\vert
^{2} \left(1 + \frac{m}{M} \right)} \, ,
\end{equation}
%
\begin{equation}
b_{\tau 0} = - \frac{2}{M \beta_{0}} a_{\tau 0} \left( 1 + \frac{m}{M}
\right)^{-1} \, ,
\end{equation}
%

\begin{equation}
b_{\tau 1} = \frac{a_{\tau 0}}{5} \left( 1 + \frac{m}{M} \right)^{-1} \, ,
\label{eq117}
\end{equation}
%
\begin{equation}
a_{L 1} = - \frac{1}{5\kappa} \sqrt{\frac{2M \beta_{0}}{\pi}} a_{\tau 0} \, .
\label{eq118}
\end{equation}

\noindent \textbf{Conflicts of interest}\\

The authors have no competing interests to declare.\\

\noindent \textbf{Declarations of interest}\\

None.\\

\noindent \textbf{Supplementary data}\\

Supplementary data to this article can be found online at https://
doi.org/.



\begin{thebibliography}{99}

\bibitem{1} Y.L. Klimontovich, A unified approach to
kinetic description of processes in active systems, in: Statistical
Theory of Open Systems, vol. 1, Kluwer Academic, Dordrecht, The
Netherlands, 1995.

\bibitem{2} S.K. Betyaev, Hydrodynamics: problems and paradoxes,
Physics-Uspekhi 38 (3) (1995) 287.

\bibitem{3} D.N. Zubarev, V.G. Morozov, I.P. Omelyan, M.V.
Tokarchuk, Unification of the kinetic and hydrodynamic approaches in
the theory of dense gases and liquids, Theor. Math. Phys. 96 (3)
(1994) 997.

\bibitem{4} B.J. Adler, D.J. Tildesley, Computer
Simulation in Liquids, Oxford Univ. Press, Oxford, UK, 1987.

\bibitem{5} M.H. Kalos, P.A. Whitlock, Monte Carlo
Methods, Wiley Interscience, New York, USA, 2007.

\bibitem{6} D.N. Zubarev, Nonequilibrium Statistical
Thermodynamics, Plenum-Consultants Bureau, New York, USA, 1974.

\bibitem{7} J.P. Dougherty, Research article foundations of
non-equilibrium statistical mechanics, Phil. Trans. Roy. Soc.
(London) A \textbf{346} (1680) (1994) 259.

\bibitem{8} R. Luzzi, A.R. Vasconcellos, J.G. Ramos,
Predictive Statistical Mechanics: A Nonequilibrium Ensemble
Formalism, Kluwer Academic, Dordrecht, The Netherlands, 2002.

\bibitem{9} R. Luzzi, A.R. Vasconcellos, J.G. Ramos, The theory
of irreversible processes: Foundations of a non-equilibrium
statistical ensemble formalism, Rivista Nuovo Cimento 29 (2) (2006)
1-85.

\bibitem{10} D.N. Zubarev, V.G. Morosov, G. R\"{o}pke, Statistical
Mechanics of Nonequilibrium Processes, vol. 1: Basic Concepts
Kinetic Theory, vol 2: Relaxation and Hydrodynamics Processes,
Akademie Verlag - Wiley VCH, Berlin, Germany, 1996.

\bibitem{11} L. Sklar, Physics and Chance: Philosophical Issuee in the
Foundations Statistical Mechanics, Cambridge Univ. Press, Cambridge,
UK, 1993.

\bibitem{12} R. Luzzi, A.R. Vasconcellos, J.G. Ramos, C.G.
Rodrigues, Statistical irreversible thermodynamics in the framework
of Zubarev's nonequilibrium statistical operator method, Theor.
Math. Phys. 194 (1) (2018) 4.

\bibitem{13} A.L. Kuzemsky, Nonequilibrium statistical operator method and
generalized kinetic equations, Theor. Math. Phys. 194 (1) (2018) 30.

\bibitem{14} V.G. Morozov, Memory effects and nonequilibrium correlations
in the dynamics of open quantum systems, Theor. Math. Phys. 194 (1)
(2018) 105.

\bibitem{15} G. R\"{o}pke, Electrical conductivity of charged particle
systems and Zubarev's nonequilibrium statistical operator method,
Theor. Math. Phys. 194 (1) (2018) 74.

\bibitem{16} H.J. Kreuzer, Nonequilibrium Thermodynamics
and its Statistical Foundations, Claredon, Oxford, UK, 1981.

\bibitem{17} H.G.B. Casimir, On Onsager's principle of microscopic
reversibility, Rev. Mod. Phys. 17 (2-3) (1945) 343.

\bibitem{18} D. Jou, J. Casas-Vazquez, G. Lebon,
Extended Irreversible Thermodynamics, Springer, Berlin, Germany,
fourth enlarged edition, 2010.

\bibitem{19} I. M\"{u}ller, T. Ruggeri, Extended
Thermodynamics, Springer, Berlin, Germany, 1993.

\bibitem{20} D. Jou, J. Casas-Vazquez, G. Lebon, Extended irreversible
thermodynamics revisited (1988-98), Rep. Prog. Phys. 62 (7) (1999)
1035.

\bibitem{21} G. Lebon, D. Jou, Early history of extended irreversible
thermodynamics (1953–1983): An exploration beyond local equilibrium
and classical transport theory, Eur. Phys. J. H. 40 (2) (2015)
205-240.

\bibitem{22} J.P. Boon, S. Yip, Molecular Hydrodynamics, McGraw-Hill,
New York, USA, 1980; Reprinted by Dover, New York, USA, 1991.

\bibitem{23} T. Dedeurwaerdere, J. Casas-V\'{a}zquez, D. Jou, G. Lebon,
Foundations and applications of a mesoscopic thermodynamic theory of
fast phenomena, Phys. Rev. E 53 (1) (1996) 498.

\bibitem{24} C.A.B. Silva, C.G. Rodrigues, J.G. Ramos, R. Luzzi,
Higher-order generalized hydrodynamics: Foundations within a
nonequilibrium statistical ensemble formalism, Phys. Rev. E 91
(2015) 063011.

\bibitem{25} C.A.B. Silva, J.G. Ramos, A.R. Vasconcellos, R. Luzzi,
Generalized kinetic equations for far-from-equilibrium many-body
systems, J. Stat. Phys. 143 (5) (2011) 1020-1034.

\bibitem{26} J.G. Ramos, C.G. Rodrigues, C.A.B. Silva, R. Luzzi, Statistical
mesoscopic hydro-thermodynamics: the description of kinetics and
hydrodynamics of nonequilibrium processes in single liquids, Braz.
J. Phys. 49 (2019) 277.

\bibitem{27} U. Fano, Description of states in quantum mechanics by density
matrix and operator techniques, Rev. Mod. Phys. 29 (1) (1957) 74.

\bibitem{28} R.P. Feynman, Lectures in Statistical Mechanics, Benjamin,
Reading, MA, USA, 1972.

\bibitem{29} R. Luzzi, A.R. Vasconcellos, J.G. Ramos,
Statistical Foundations of Irreversible Thermodynamics,
Teubner-BertelsmannSpringer, Stuttgart, Germany, 2000.

\bibitem{30} R. Luzzi, A.R. Vasconcellos, J.G. Ramos, Non-equilibrium statistical
mechanics of complex systems: An overview, Rivista Nuovo Cimento 30
(3) (2007) 95-157.

\bibitem{31} R. Luzzi, A.R. Vasconcellos, Ultrafast
Transient Response of nonequilibrium plasma in semiconductors, in:
Semiconductor Processes Probed by Ultrafast Laser Spectroscopy, vol.
1, edited by R. R. Alfano, Academic, New York, USA, 1984.

\bibitem{32} C.G. Rodrigues, V.N. Freire, J.A.P. Costa, A.R.
Vasconcellos, R. Luzzi, Hot electron dynamics in zincblende and
wurtzite GaN, Phys. Stat. Sol. (b) 216 (1999) 35.

\bibitem{33} C.G. Rodrigues, A.R. Vasconcellos, R. Luzzi, Ultrafast
relaxation kinetics of photoinjected plasma in III-nitrides, J.
Phys. D 38 (2005) 3584.

\bibitem{34} C.G. Rodrigues, A.R. Vasconcellos, R. Luzzi, Non-Linear
electron mobility in n-doped III-Nitrides, Braz. J. Phys. 36 (2006)
255.

\bibitem{35} A. Ott, J.P. Bouchaud, D. Langevin, W. Urbach, Anomalous
diffusion in ``living polymers": A genuine Levy flight?, Phys. Rev.
Letters 65 (1990) 2201.

\bibitem{36} A.I. Akhiezer, S.V. Peletminskii, Methods of Statistical
Physics, Pergamon, Oxford, UK, 1981.

\bibitem{37} L. Lauck, A.R. Vasconcellos, R. Luzzi, A nonlinear quantum transport
theory, Physica A 168 (1990) 789-819.

\bibitem{38} J.R. Madureira, A.R. Vasconcellos, R. Luzzi, L.
Lauck, Markovian kinetic equations in a nonequilibrium statistical
ensemble formalism, Phys. Rev. E 57 (1998) 3637.

\bibitem{39} J.C. Maxwell, On the dynamical theory of gases, Phil. Trans.
Roy. Soc. (London) 157 (1867) 49-88.

\bibitem{40} C.G. Rodrigues, C.A.B. Silva, J.G. Ramos, R. Luzzi,
Maxwell times in higher-order generalized hydrodynamics: Classical
fluids, and carriers and phonons in semiconductors, Phys. Rev. E 95
(2017) 022104.

\bibitem{41} R. Balian, Y. Alhassed, H. Reinhardt, Dissipation
in many-body systems: A geometric approach based on information
theory, Phys. Rep. 131 (1-2) (1986) 1.

\bibitem{42} J.G. Ramos, A.R. Vasconcellos, R. Luzzi, A nonequilibrium
ensemble formalism: Criterion for truncation of description, J.
Chem. Phys. 112 (6) (2000) 2692.

\bibitem{43} R. Luzzi, M.A. Scarparo, J.G. Ramos, A.R.
Vasconcellos, M.L. Barros, Z. Zhiayao, A. Kiel, Informational
statistical thermodynamics and thermal laser stereolithography, J.
Non-Equil. Thermodynamic. 22 (3) (1997) 197-216.

\bibitem{44} C.G. Rodrigues, A.R.B. Castro, R. Luzzi, Higher-order
generalized hydrodynamics of carriers and phonons in semiconductors
in the presence of electric fields: Macro to nano, Phys. Stat.
Solidi B 252 (2015) 2802.

\bibitem{45} A.R. Vasconcellos, A.R.B. Castro, C.A.B. Silva, R.
Luzzi, Mesoscopic hydro-thermodynamics of phonons, AIP-Advances 3
(2013) 072106.

\bibitem{46} C.G. Rodrigues, A.R. Vasconcellos, R. Luzzi, Mesoscopic
hydro-thermodynamics of phonons in semiconductors: heat transport in
III-nitrides, Eur. Phys. J. B 86 (5) (2013) 200.

\bibitem{47} C.G. Rodrigues, A.R. Vasconcellos, R. Luzzi, Thermal
conductivity in higher-order generalized hydrodynamics:
Characterization of nanowires of silicon and gallium nitride,
Physica E 60 (2014) 50.

\bibitem{48} L.D. Landau, E.M. Lifschitz, Theory of
Elasticity, Pergamon, Oxford, UK, 1986.

\bibitem{49} R. Balian, From Microphysics to Macrophysics, vol. 2, Springer,
Berlin, Germany, 2007.

\bibitem{50} O.S. Duarte, A.O. Caldeira, Effective coupling between
two brownian particles, Phy. Rev. Letters 97 (2006) 250601.

\bibitem{51} A.F. Fonseca, M.V. Mesquita, A.R. Vasconcellos, R.
Luzzi, Informational-statistical thermodynamics of a complex system,
J. Chem. Physics 112 (2000) 3967-3979.

\bibitem{52} M.V. Mesquita, A.R. Vasconcellos, R. Luzzi, Amplification
of coherent polar vibrations in biopolymers: Fr\"{o}hlich
condensate, Phys. Rev. E 48 (5) (1993) 4049.

\bibitem{53} C.G. Rodrigues, F.S. Vannucchi, R. Luzzi, Non-equilibrium
Bose-Einstein-like condensation, Advanced Quantum Technologies 1
(2018) 201800023.

\bibitem{54} C.G. Rodrigues, A.R. Vasconcellos, R. Luzzi, Evolution
kinetics of nonequilibrium longitudinal-optical phonons generated by
drifting electrons in III-nitrides: longitudinal-optical-phonon
resonance, J. Appl. Phys. 108 (2010) 033716.

\bibitem{55} C.G. Rodrigues, A.R. Vasconcellos, R. Luzzi, Nonlinear
transport in n-III-nitrides: Selective amplification and emission of
coherent \textsc{lo} phonons, Solid State Commun. 140 (2006) 135.

\bibitem{56} C.G. Rodrigues, A.R. Vasconcellos, R. Luzzi, Drifting electron
excitation of acoustic phonons: Cerenkov-like effect in n-GaN, J.
Appl. Phys. 113 (2013) 113701.

\bibitem{57} R. Balescu, Equilibrium and Nonequilibrium Statistical
Mechanics, Wiley-Interscience, New York, USA, 1975.


\end{thebibliography}
\end{document}